# Analyzing the response to TV serials retelecast during COVID19 lockdown in India

**Sandeep Ranjan**

**Associate Professor, AIMETC, Jalandhar, INDIA**

**sandeep.ranjan@learn.apeejay.edu**

**Abstract:**

TV serials are a popular source of entertainment. The ongoing COVID19 lockdown has a high probability of degrading the public's mental health. The Government of India started the retelecast of yesteryears popular TV serials on public broadcaster Doordarshan from 28th March 2020 to 31st July 2020. Tweets corresponding to the Doordarshan hashtag were mined to create a dataset. The experiment aims to analyze the public's response to the retelecast of TV serials by calculating the sentiment score of the tweet dataset. Dataset's mean sentiment score of 0.65 and high share (64.58%) of positive tweets signifies the acceptance of Doordarshan's retelecast decision. The sentiment analysis result also reflects the positive state of mind of the public.

## Introduction

The ongoing novel coronavirus (COVID19) pandemic has affected human life and behavior since its outbreak in December 2019 (Dubey, 2020). Humanity is fighting for its survival by observing lockdowns in almost every country. The curve of the spread of the disease is very steep with the cases doubling at a very fast pace. Leaders of different countries called for lockdowns after analyzing the situations and making necessary arrangements. In India, the first lockdown of 21 days was called from 24th March 2020 (Pulla, 2020). People had been confined to their homes and all outdoor activities except emergencies ceased.

COVID19 vaccine will be available to the public in about 18 months and may still not be effective in bringing life back to normal (Adalja et al., 2020). Health workers and researchers don't have clear predictions about the spread of the pandemic because of asymptomatic and symptomatic case data in different countries. COVID19 is going to bring tough times for the general public, healthcare workers and healthcare infrastructure. Social distancing and "work from home" can prove effective in preventing the spread of the pandemic.

Fear and anger are the most visible emotions observed in critical times of pandemics like the COVID19 (Ornell et al., 2020). The cases of mental health degradation outnumber the cases of physical health degradation during pandemics due to the spread of sentiments and information. Media and entertainment platforms have been used to boost the morale of the general public in various countries (Parivudhiphongs, 2020). Apartment singing, mobile light flashing, clapping, YouTube and TikTok platforms have encouraged the people to stay cheerful and fight the pandemic with full enthusiasm. These activities and events boost the morale of the people by giving them a sense of being a part of the community where everyone is united and working hard to achieve the common goal- back to normal life.

The Indian Prime Minister, Mr. Narendra Modi while addressing the public during the COVID19 lockdown talked about the ancient Hindu scriptures "Ramayana and Mahabharta" and advised everyone to follow the highest virtues and self-control shown by the characters of the sacred scripts (Bansal, 2020). Referring to the novel coronavirus as a daemon, he asked people to observe social distancing to gain victory on it. Subsequent retelecast of the TV serials based on the sacred scriptures gained huge popularity in the lockdown times.

The national TV broadcaster of India, *Doordarshan*, under the aegis of the Ministry of Broadcasting, India has taken a lead in keeping the people engaged in tough times of COVID19 lockdown (Verma et al., 2020). Doordarshan is retelecasting its all-time popular TV serials and gained high Television Rating Point (TRP) for retelecasted serials (*Sunam, 2020*.). On 13th April 2020, Doordarshan launched a new TV channel "DD Retro" which is dedicatedly airing yesteryear popular TV serials (*India TV*, 2020). The presented experiment is an attempt to study the effect of Doordarshan's strategy of retelecasting old TV serials on the psychological and mental state of Indian people during the COVID19 lockdown period. The frequency of words occurring in the dataset and sentiment analysis of the entire dataset gives an insight into the response of the general public to the TV serial retelecast decision.

## Literature review

An extensive study of the literature reveals that there is a need to study the ill effects of COVID19 on the psychological health of the masses and also to plan for the mental wellbeing of people. There is always a high probability of psychological and mental health degradation in situations like COVID19 lockdown (Brooks et al., 2020). Stress may arise out of fear of infection, financial loss, isolation, disease, grocery supplies. A sudden lockdown caused by a pandemic like the COVID19 can have a significant effect on the mental health of people (Rajkumar, 2020). Psychiatrists are analyzing the conversations of people to trace negativity or positivity in them for establishing a correlation with the mental health of respondents.

People now restricted to their homes are showing increased engagement and activity on social media platforms (Cinelli et al., 2020). The lightning-fast rate of information dispersal on social networks continues to improve with each passing day. The study was conducted on COVID19 related popular keywords trending on Google Trends in the first half of January 2019. Social networks started getting the attention of the public as soon as the news related to the pandemic broke out and people vent out their emotions on the online platforms expressing concerns, preparedness, fear. Cross-platform sharing of the information fueled the spread of pandemic related content in a very short period.

Many countries have initiated measures for dealing with the psychological health of the public in crisis times (Duan & Zhu, 2020). Focused studies have revealed that the pandemic has caused significant degradation of the mental health of the public in general and health workers in particular. These studies emphasize the need for actions and interventions to tackle the ill effects on the psychological state of mind of those affected.

The Indian Prime Minister has carried out several activities to boost the morale of the citizens in general and emergency service providers in particular (Vibha et al., 2020). Twitter was mined for the popular hashtags #IndiaLockdown and #IndiafightsCorona in the initial days of the first lockdown in India. Despite the prevailing fear and negativity, the overall sentiment of the general public was positive. Analyzing the tweets related to the hashtags for such activities reveals that the general public has shown support for the actions of the government and reflects a positive behavior.

The Indian public has given a positive response to the initial 21 days of lockdown announced by the Indian Government (Kaur & Ranjan, 2020). Top words in the tweet dataset correspond to the emotions of the general public about the shortage of groceries and fear related to it. Policymakers can gauge the effect of COIVD19 lockdown by mining tweets related to the situation and get real-time feedback on the corrective actions being taken by them. The analysis of the tweets for the 21 day lockdown period revealed that the overall sentiment was negative only for 3 days compared to the positive sentiment for 18 days.

Television acts as a crucial link between government and people during crisis times (Ayedee & Manocha, 2020). TV has emerged as a major source of information dissemination and entertainment for the general public. In a densely populated country like India where things may worsen due to COVID19, television helps to maintain peace and a calm atmosphere. Policymakers have extensively used television and other sources of information dissemination to plan and execute mental health care during COVID19 (Ransing et al., 2020). Researchers developed a framework to assess the general public's mental health status in response to the information shared by the government using various media. Massive datasets have been created from social networks to analyze the mental health of the public to develop strategies in the fight against COVID19.

**Research objective**

The main objective of this experiment is to assess the general public's response to India's national TV broadcaster Doordarshan's decision to retelecast popular TV serials during COVID19 lockdown by mining hashtag tweets.

**Research methodology**

In times of pandemics, governments must engage in communication with the general public to contain fear and panic (Chen et al., 2020). The latest machine learning techniques have been utilized to monitor the psychological well-being of the masses (Calvo et al., 2020). Analyzing well being of people can help policymakers monitor the crisis and thus plan accordingly. Online social networks provide crucial and timely information about the sentiments of users (Song et al., 2019). Social networks serve as inexpensive means of easy and speedy communication (Petrilák et al., 2020). Indian public via social media platforms had demanded the Government of India to retelecast old TV serials and honoring the demands of the Indian public, the Minister of Information and Broadcasting, Government of India announced the retelecast of popular Indian TV serial "Ramayan" from 28th March 2020 (*OutlookIndia, 2020*). Since then many old TV serials including "Mahabharat", "Sri Krishna", "Vishnu Puran", "Chanakya", "Buniyaad" and "Shaktiman" have been retelecasted on various TV channels of Doordarshan (*Sacnilk*, 2020). The presented experiment started with mining tweets for hashtags related to the national TV broadcaster Doordarshan from 27th March 2020, a day before the first retelecast of TV serial "Ramayana" Tweets were mined till 31st July 2020 when India had already prepared itself to unlock the lockdown caused by COVID19.

Tweepy, a Python library was used to mine tweets for the hashtag #doordarshan and another Python library, TextBlob was used to perform sentiment analysis on the mined tweet datasets. After cleaning the dataset for noise and duplicate tweets, the number of tweets for further processing is 7669. Figure 1 shows the snapshot of the tweet dataset mined for the experiment. Table 1 shows the list of most frequently used words and word pairs in the dataset.

**Figure 1.** A snapshot of the tweet dataset

| Relationship Date (UTC) | Tweet |
|---|---|
| 14-05-20 7:37 | RT @MumbaiMirror: Doordarshan is all set to air @iamsrk's small screen classics #DoosraKeval.https://t.co/RagOjCRc6k |
| 14-05-20 7:39 | Shah Rukh Khan's lesser known show Doosra Keval to re-air on Doordarshan https://t.co/SexPJapjIt |
| 14-05-20 7:43 | @StarSportsIndia @msdhoni Doordarshan must telecast finals of wc.... Of the legend kapil dev n dhoni.... |
| 14-05-20 7:46 | #NitishBharadwaj starrer #VishnuPuran makes a comebackhttps://t.co/ndZgRD4qJt |
| 14-05-20 7:48 | After Mahabharat, Doordarshan all set to bring Nitish Bharadwaj starrer Vishnupuranhttps://t.co/7zw4r2UkYd |
| 14-05-20 7:52 | #buniyaad serial from the doordarshan era is bringing found memories for so many of us...#buniyaad... https://t.co/OIGZaIf7VX |
| 14-05-20 7:52 | RT @ZeeNewsEnglish: After 'Circus', 'Fauji', #ShahRukhKhan's #DoosraKeval TV show makes a comeback on #Doordarshanhttps://t.co/cgS52OWq7l |
| 14-05-20 7:59 | RT @Bollyhungama: .@iamsrk's lesser-known show #DoosraKeval to re-air on Doordarshanhttps://t.co/8SEipPfB4l |
| 14-05-20 8:02 | @DDNational Now it is the time to telecast the heroic stories of India's brave son Maharana Pratap Doordarshan!!!... https://t.co/4FILZDDEac |
| 14-05-20 8:07 | We welcome Dr Vijaylaxmi Chhabra, Former Director General Doordarshan, who'll talking to our students and staff on... https://t.co/dUGKlXjIPv |
| 14-05-20 8:08 | RT @gulftoday: #COVID－19 : Shah Rukh Khan's show from 1989, 'Doosra Keval,' to be rerun on Doordarshan https://t.co/xil0NW4FDa |
| 14-05-20 8:08 | RT @TellyTalkIndia: #NitishBharadwaj starrer #VishnuPuran makes a comebackhttps://t.co/ndZgRD4qJt |

**Source:** Author's dataset created for the research

**Table 1.** Most frequently used words and word pairs

| Word | Frequency | Word pairs | Frequency |
|---|---|---|---|
| doordarshan | 1157 | doordarshan, ramayan | 1550 |
| ddnational | 168 | flop, show | 1494 |
| show | 137 | ddnational, doordarshan | 1469 |
| ramayan | 137 | shrikrishna, ddnational | 1463 |
| mahabharat | 97 | ramayan, prasarbharti | 1112 |
| watch | 97 | prasarbharti, ramayanonstarplus | 1024 |
| news | 96 | ramayanonstarplus, ramayanonddnational | 733 |
| india | 94 | return, doordarshan | 642 |
| hai | 88 | doordarshan, channel | 460 |
| channel | 84 | ramayan, mahabharat | 420 |

**Source:** Author's dataset created for the research

Figure 2 depicts the word cloud of the most frequently used words in the dataset. Word clouds sum up social networks as the collective wisdom of the masses as unrelated individuals participate in topics of common interest (Wang, 2019). The collective wisdom of the crowds can help assess a particular situation or predict its outcomes (Bari et al., 2019). The word cloud presented in the figure has been created on the map of India representing sentiments of the Indian public in the form of common words used in the posts shared for the Twitter handles about Doordarshan and the TV serials being retelecasted during the COVID19 lockdown period. Words having the highest frequency of occurrence in the dataset are the biggest and brightest in the word cloud picture. A high concentration of positive sentiment words in the dataset makes the word cloud a reflection of a positive state of mind of the general public.

**Figure 2.** Word cloud of most frequently used words

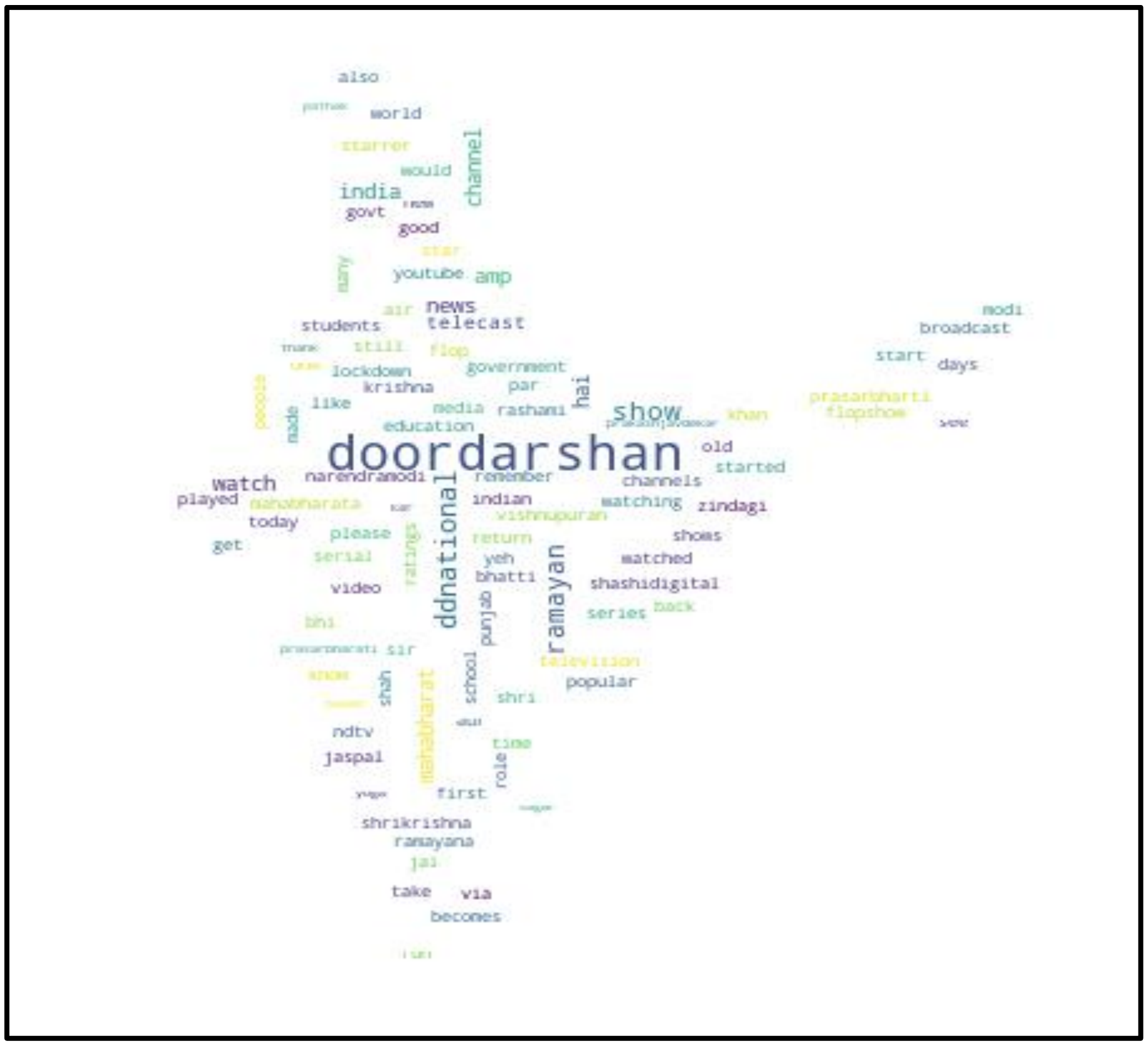

**Source:** Word cloud generated for the experiment's dataset using Python

Each tweet in the dataset has been categorized as positive, negative or neutral based on the sentiment conveyed in it. The sentiment value ranges from -1 to +1 based on the content of the tweets and corresponding subjectivity. The mean sentiment score of the tweet dataset is 0.65. 4953 tweets (64.58%) turned out to be positive representing a positive state of mind towards TV serial retelecast decision. 841 neutral tweets contain phrases and words whose sentiment/ meaning could not be generated. Often social network mined datasets contain a large number of text items that can not be associated with any particular entity or event. Figure 3 shows the pie chart of the number of positive, negative and negative tweets. 1875 negative tweets contain the words for fear, anger, panic and shortage of supplies and money.

**Figure 3.** Tweet categorization

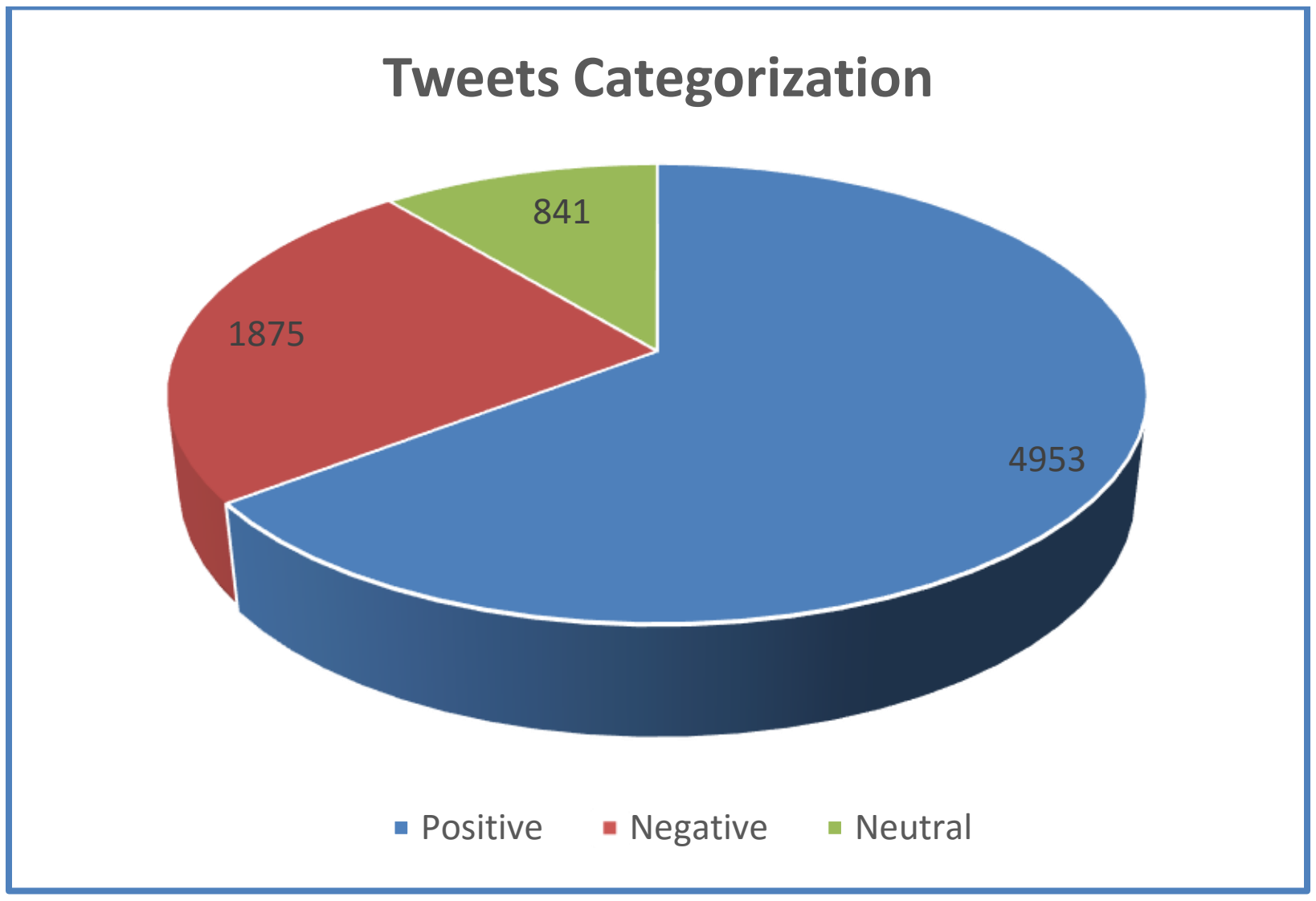

**Source:** sentiment analysis of the experiment's dataset

Analyzing the dataset reveals people are happy with the decision of the Government of India to retelecast popular TV serials of yesteryears. Apart from sharing opinions on their favorite TV serials and Doordarshan, many users have requested to retelecast some other TV serials, historical cricket matches and hockey matches played by the Indian teams. The positive sentiments towards Doordarshan and its popular TV serials like Ramayan reflect in the record TRP of 170 million viewers in the first week of April 2020 (*Republic World*, 2020). The TV serial retelecast decision has a positive effect on the general public in realigning their daily routine during the COVID19 lockdown (Tripathi et al., 2020). People have reset their routine and started watching the episodes of the TV serials punctually. Inactivity or lack of physical activity impacts the physical and medical health of people (Romero-Blanco et al., 2020). An active and punctual lifestyle promotes good health. The tweets in the dataset thank Doordarshan for preventing their daily routine to become sedentary.

## Results & conclusion

Doordarshan which had been the numero uno entertainer since its inception till the advent of private TV channels has made a strong comeback in the Indian scenario (*Deccan Herald, 2020*). The new "work from home" culture has caused parents to spend more time with their kids and watch TV serials with them. A large number of requests for retelecasting TV serials /sports coverage and a mean sentiment score of 0.65 confirms the support towards Doordarshan's decision and its positive impact on the public's mental health. TV serials like Ramayana made it to the most frequently mentioned words in the mined dataset and "Doordarshan, Ramayana" is the most frequently mentioned word pair.

The mean sentiment score of the dataset is 0.65 which is quite positive (sentiment ranges from -1 to +1) during the lockdown times. The retelecast of favorite TV serials by the national broadcaster has a positive effect on the sentiments and psychiatric health of Indian people. Only 1875 tweets out of the total 7669 tweets (24.45% share) were found to be negative. 4953 tweets were positive (64.58% share) and 841 tweets were neutral (for which the sentiment could not be determined) signify that the public is highly positive in their opinions.

## Research implications

The experiment aims to assess the public's response to the TV serial retelecast decision during the COVID19 pandemic. The results of this experiment can be of great use to the policymakers to validate their decision to use the national broadcaster Doordarshan's popular serials in improving the general public's mental health. A high positive mean score of the dataset having 64.58% positive tweets for the study period signifies the success of TV serial retelecast. The limitations of the experiment need to be considered to prevent the over-generalization of results.

## Research Limitation:

The presented experiment is limited to the tweet dataset of hashtags related to Doordarshan and its most popular TV serials that are being retelecasted during the COVID19 pandemic. More than 20 such serials are being retelecasted and the number has been rising owing to the popularity of retelecasted serials. The experiment can be improved by including all retelecasted serials and the public's response to them on major social networks.